\documentclass[aps,prb,twocolumn,superscriptaddress,showpacs]{revtex4-1}

\usepackage{graphicx}
\usepackage{dcolumn}
\usepackage{amsmath,bm}
\usepackage{color}%

\begin{document}

\title{Interlayer coupling enhancement in graphene/hexagonal boron nitride heterostructures by intercalated defects or vacancies}

\author{Sohee Park}
\affiliation{Department of Materials Science and Engineering,
             Seoul National University, Seoul, 151-747, Korea}

\author{Changwon Park}
\affiliation{Center for Nanophase Materials Sciences,
             Oak Ridge National Laboratory, Oak Ridge, Tennessee 37831, United States}

\author{Gunn Kim}
\email[Corresponding author: ]{gunnkim@sejong.ac.kr}
\affiliation{Department of Physics and
             Graphene Research Institute,
             Sejong University, Seoul, 143-747, Korea}

\date{\today}

\begin{abstract}
Hexagonal boron nitride (hBN), a remarkable material with a two-dimensional atomic crystal structure, has the potential to fabricate heterostructures with unusual properties.
We perform first-principles calculations to determine whether intercalated metal atoms and vacancies can mediate interfacial coupling
and influence the structural and electronic properties of the graphene/hBN heterostructure.
Metal impurity atoms (Li, K, Cr, Mn, Co, and Cu), acting as extrinsic defects between the graphene and hBN sheets, produce $n$-doped graphene.
We also consider intrinsic vacancy defects and find that a boron monovacancy in hBN acts as a magnetic dopant for graphene,
whereas a nitrogen monovacancy in hBN serves as a nonmagnetic dopant for graphene.
In contrast, the smallest triangular vacancy defects in hBN are unlikely to result in significant changes in the electronic transport of graphene.
Our findings reveal that an hBN layer with some vacancies or metal impurities enhances the interlayer coupling in the graphene/hBN heterostructure
with respect to charge doping and electron scattering.
\end{abstract}


\maketitle

\section{Introduction}

Silica (SiO$_2$) is the most commonly used substrate for
graphene-based devices, and while it offers many advantages, there
are also several drawbacks. Graphene supported on SiO$_2$
exhibits an electron mobility that is 10 times lower than that of the structure suspended in vacuum\cite{Geim,Neto},
owing to the interfacial charge impurities,\cite{Chen,Hwang} surface
roughness,\cite{Ishigami,Morozov,Stolyarova} and surface optical
phonons\cite{Chen,Fratini} of the silica substrate. Further, the roughness of the silica
surface results in nanometer-scale ripples\cite{Ishigami,Stolyarova,Ryu} in graphene.

Hexagonal boron nitride (hBN) is a III-V compound similar to
graphite and consists of equal numbers of boron and nitrogen atoms;
it is a chemically and thermally stable dielectric material with an
optical band gap of $\sim$6 eV.\cite{Watanabe}
Since graphene devices on hBN have a much higher carrier
mobility\cite{Dean} than devices on SiO$_2$, the lack of surface
dangling bonds, small roughness, and slight lattice
mismatch\cite{Giovannetti} with graphene mean that high-quality
graphene devices could be realized using a few-layer hBN film as an
insulating substrate.

Monolayer hBN, also known as ``white graphene,'' can be peeled off
from bulk hBN by mechanical cleavage,\cite{Dean,Pacile,Lee} although chemical vapor deposition (CVD) is
generally used to produce large-area hBN sheets.\cite{Shi, Kim} Recently, a graphene
monolayer was directly grown on a CVD-grown hBN film\cite{MWang}; 
the monololayer grown in this manner enabled superior electronic properties as compared to those in the case of graphene
transferred to the hBN film. However, growing hBN sheets via CVD
leads to the introduction of native defects.\cite{CJin,VACCVD}

More recently, van der Waals (vdW) heterostructures, comprising graphene and other two-dimensional atomic crystals such as hBN, MoS$_2$, and WS$_2$, 
have been produced and examined in this emerging research field of condensed matter physics.\cite{Geim2013}
These heterostructures are expected to show new physics.
{Although the vdW interaction is much weaker than covalent bonding,
it is sufficiently strong to maintain the stacking of two-dimensional atomic crystals.
Thin hBN layers can serve, not only as good substrates, but also as protective covers for graphene.\cite{Mayorov}
Another attractive topic is the vertical devices or tunneling heterostructures
composed of graphene and few-layer-thick crystals of hBN or transition metal dichalcogenide.\cite{Britnell, Georgiou}
These devices constitute a new type of electronic devices.}
The interfacial coupling in the vdW heterostructures could be enhanced through different phenomena, 
for instance, interfacial contamination by adsorbates and native defects \cite{Geim2013}.
At such a contaminated interface, localized gap states
originating from the vacancies or intercalated metal atoms
could cause electronic scattering near the Fermi level ($E_f$) of the graphene sheet.

In this paper, we report a first-principles study of the structural and electronic properties of a graphene/hBN heterostructure containing defects.
We consider metal impurity atoms on the hBN sheet as extrinsic defects and vacancies in hBN as
intrinsic defects. The electronic structure is calculated to determine the energies of the defect states; 
the results show that the interlayer coupling is enhanced in the presence of the defects. 
The defect-mediated interfacial coupling leads to residual electronic scattering in graphene/hBN heterostructures 
containing defects such as vacancies and adsorbed impurity atoms. 
This in turn may be the origin of charge doping and residual electronic scattering in graphene.

\begin{figure}[tb!]
\includegraphics[width=0.7\columnwidth]{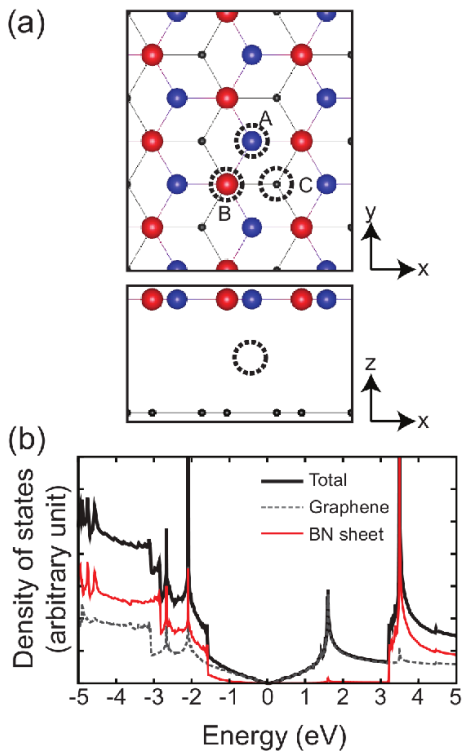}
\caption{(Color online) (a) Atomic structure of a graphene/hBN heterostructure
and (b) the density of states of our model structure.
Black, red, and blue spheres denote carbon, boron, and nitrogen
atoms, respectively, and the dotted circle represents the position
of the intercalated metal atom.} \label{Fig1}
\end{figure}

\section{Computational details}

We adopted the vdW-corrected density functional theory (DFT)
calculations of the Vienna ab-initio simulation package
(VASP)\cite{VASP} to investigate the structural, electronic, and magnetic
properties of a graphene/hBN heterostructure.
To examine the effects of an impurity metal atom, we included a Li, K,
Cr, Mn, Co, or Cu atom between the hBN and graphene sheets in our model.
To examine the Coulomb interaction and/or the orbital hybridization of the intercalated atom,
we introduced metal atoms (Li, K, Cr, Mn, Co, and Cu).
The ionic or covalent bonding can enhance the interlayer coupling between graphene and hBN.
In particular, an alkali metal, such as Li and K, can easily donate an electron to the sheets.
We compared the Li and K atoms with regard to the dependence of the metal atom radius on the deformation and the amount of donated charge. 
Because of the presence of {\it d} orbitals, 3{\it d} transition metal atoms may have strong coupling to the graphene and hBN layers.
In addition, as the transition metal atoms may cause a spin magnetic moment, we considered Cr, Mn, and Co atoms as magnetic dopants.
Copper surfaces have been used for the CVD growth of graphene and hBN sheets.
If the Cu atoms or nanoclusters are not completely removed from the hBN or graphene sheet
after the growth, the graphene/hBN heterostructure may contain copper impurity atom(s) at the interface.
Thus, we also considered Cu as an impurity.
During the CVD growth of the hBN sheet, monovacancies and
triangular defect structures such as $V_{B+3N}$ and $V_{N+3B}$
defects can be formed\cite{VACCVD} [$V_{B+3N}$ ($V_{N+3B}$) is a
multi-vacancy in which one nitrogen (boron) and three boron (nitrogen)
atoms are missing]. Thus, we considered four different vacancy
structures in hBN: a single boron vacancy ($V_B$); a single nitrogen
vacancy ($V_N$); and the two triangular vacancies, $V_{N+3B}$ and
$V_{B+3N}$.

Our supercell was an orthorhombic cell containing a graphene sheet and an hBN sheet.
The cell size was 17.18 $\times$ 14.96 $\times$ 25.00 \AA$^3$,
with the vacuum region being about 22 \AA~thick. 
All the model systems were relaxed until the residual atomic forces were less than 0.02 eV$\cdot$\AA$^{-1}$.
To describe the interaction between the electrons and ions, we used the
generalized gradient approximation (GGA) with a functional form proposed in Ref.~\cite{PBE}
within the projector-augmented wave (PAW) method.\cite{PAW}
An energy cutoff of 400 eV and a $\Gamma$-centered 3 $\times$ 3 $\times$ 1
$k$-point sampling were employed to calculate the total energy and
to obtain fully relaxed geometries. We used a grid of
6 $\times$ 6 $\times$ 1 $k$-points to plot the density of states and
charge densities. The vdW interaction was also considered
to precisely describe the structural properties between the two
layers.
In the calculations, we made a correction of the vdW interaction
between the metal impurity and the two sheets, using Grimme's DFT$+$D2 method.\cite{Grimme}
The large Coulomb repulsion of the localized $d$ and 
$f$ electrons may not be well represented by a conventional DFT
functional. To correctly describe the localization characteristics of 3$d$ orbitals, a Hubbard $U$ term was
added to the DFT functional (i.e., the GGA+$U$ method).

\section{Results and discussion}
\subsection{Extrinsic defects: intercalated metal atoms}

We consider the most stable stacking configuration among three inequivalent
configurations\cite{Giovannetti,Slawinska} in the absence of any
defects [Figure 1(a)]. The most stable structure is realized with
Bernal (AB) stacking of the hBN and graphene sheets, in a manner similar
to that of graphite, where the nitrogen atom is located above the
center of the graphene hexagons, as shown in Figure 1(a). In the
fully optimized (commensurate) configuration, the distance between the
nearest-neighbor atoms is 1.44 \AA, which is close to the average
carbon--carbon distance in graphene (1.42 \AA) and B--N distance in
the BN sheet (1.45 \AA); the distance between the graphene and hBN
sheets is 3.09 \AA. The adatoms were initially placed in the center
of the two layers (hBN and graphene) at the position indicated by
the dotted circle in Figure 1(a). The impurity atom is on top of a
nitrogen (boron) atom at site A (B), while it is at a hollow site of
the hBN layer at site C.

\begin{table}[b]
\caption{Intercalation energies (eV) of the metal-atom-sandwiched graphene/hBN structures.
\label{Table1}}%
\begin{ruledtabular}
\begin{tabular}{ccccccc}
              &  Li  & K  & Cr  & Mn &  Co  & Cu \\
\hline
     site A   &  $-2.48$   & $-0.17$ & $-0.98$ & $-1.08$ & $-2.58$ & $-1.37$ \\
     site B   &  $-1.67$  & $+0.02$ & $-0.05$  & $+0.01$  & $-1.72$ & $-0.58$ \\
     site C   &  $-1.66$  & $+0.01$ & $+0.02$  & $-0.02$ & $-1.32$ & $-0.56$ \\
\end{tabular}
\end{ruledtabular}
\end{table}

The densities of states (DOS) of our model in the absence of an
impurity metal atom is shown in Figure 1(b); the Fermi level is set to zero.
The thick solid black line represents the total DOS,
while the dashed gray and solid red lines denote the projected densities of states (PDOS) of
graphene and hBN, respectively. The overall features of the two PDOS
are similar to those of their free-standing equivalents. However, the hBN monolayer has a small DOS, between $-$1.5 and $-$1.0 eV, 
owing to the hybridization between the graphene and hBN layers. 
Thus, the band gap seems to have decreased.

\begin{figure}[tb!]
\includegraphics[width=0.85\columnwidth]{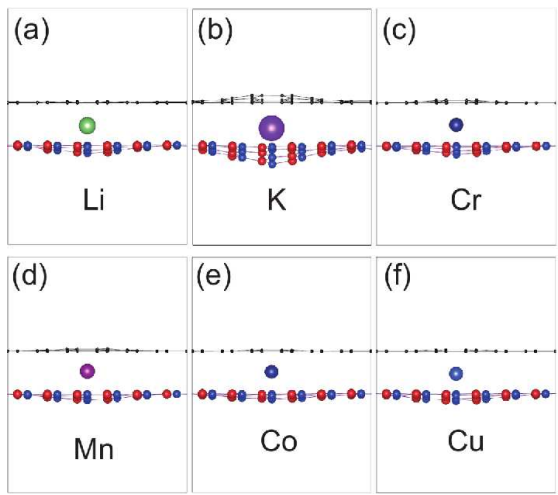}
\caption{(Color online) Optimized geometries containing (a) Li, (b)
K, (c) Cr, (d) Mn, (e) Co, and (f) Cu impurity atoms sandwiched
between the graphene and hBN sheets. \label{Fig2}}
\end{figure}

All the impurity metal atoms energetically prefer site A, which means that
the metal atoms are likely to bind to a nitrogen atom in the hBN sheet.
Table I shows the intercalation energies (in eV) of the three adsorption sites for each impurity atom, defined by
\begin{equation}
E_{\rm int} = E[{\rm graphene/{\it M}/hBN}] - E[{\rm graphene/hBN}] - E[M],
\end{equation}
where {\it M} represents the metal atom.
The negative (positive) sign represents an exothermic (endothermic) process.
For Li, the energies for intercalation are almost the same at both sites B and C. This is also $\approx$0.8 eV higher than the energy at site A.
For 3$d$ transition metals, sites B and C need 1--2 eV higher energies for intercalation than those for site A.
Although the hBN sheet is severely deformed, the structure of the graphene sheet
does not change significantly except for a K impurity atom. For K-doping, the B--N bond lengths near the impurity atom
increase by a maximum of $\sim$2.8\%. Such deformation is responsible for the lowest intercalation energy ($-0.17$ eV) for site A. 
In addition, unlike Li and K, transition metals only cause a slight change in the
graphene/hBN structure. The stronger Coulomb attraction between a
positively charged metal ion and the electron-doped graphene sheet
prevents deformation of the graphene. In contrast, the permanent dipoles of the
B--N bonds in the hBN sheet and the lower electron doping result in
a weaker attractive force between the hBN sheet and the metal ion,
leading to significant deformation of hBN.
The impurity atoms between the graphene and hBN sheets could be detected
and mapped by atomic force microscopy (AFM) and atomic force acoustic microscopy (AFAM) when such severe deformation occurs, as shown in Figure 2(b).
It is possible to observe the deformation of the graphene layer caused by the intercalated atom using AFM.
Using AFAM, which is sensitive to the tip-sample contact stiffness,
mapping of the contact stiffness could reveal a change near the impurity between graphene and hBN.

\begin{figure}[tb!]
\includegraphics[width=0.85\columnwidth]{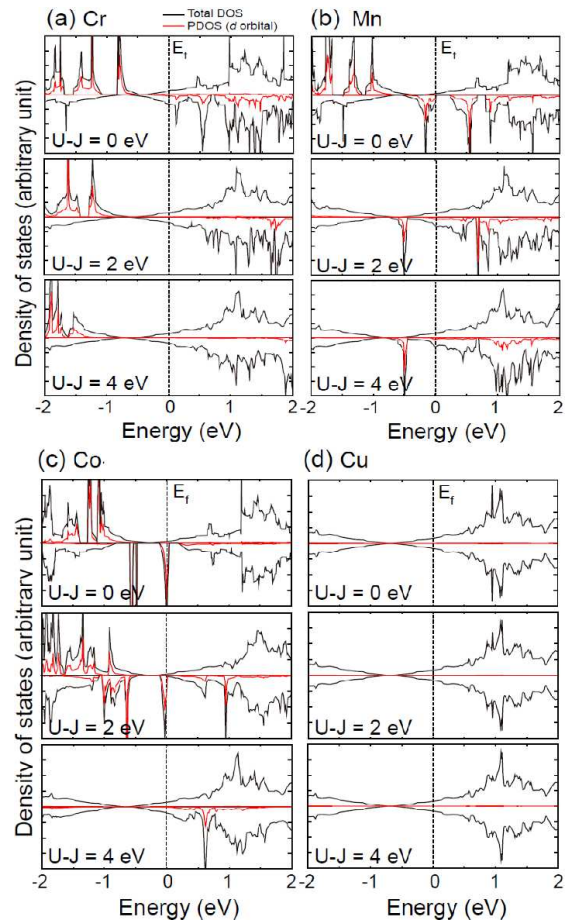}
\caption{(Color online) Projected densities of states of the
structures containing (a) Cr, (b) Mn, (c) Co, and (d) Cu impurity
atoms depending on $U-J$ of the 3$d$ orbitals. \label{Fig3}}
\end{figure}

Inserting Li or K between the two layers shifts the Fermi level to a
higher energy. According to the results of a Bader charge analysis,
almost one whole electron of the alkali impurity atom is donated to the
graphene and hBN sheets, and most of the charge prefers to move to
the graphene. For Li, 0.83 $e$ is transferred to the graphene and
0.17 $e$ to hBN. For K, on the other hand, 0.77 $e$ moves to graphene
and 0.03 $e$ to hBN. Since the impurity atom is charged, it forms a long-range Coulomb potential, which can cause weak carrier scattering.\cite{Chen2}
For 3$d$ transition metal impurities (Cr, Mn, Co, or Cu), strong orbital hybridization with the two
sheets, as well as charge transfer, play crucial roles in modifying
the electronic and magnetic properties of the graphene/metal atom/hBN sandwich structure.
In this case, a short-range potential due to the covalent bonds can have intenser effects.\cite{Suzuura1, Suzuura2}

\begin{table}[b]
 \caption{Electrons (in units of $e$) donated by the transition metal atoms and the spin magnetic moments (in units of $\mu_B$) of the graphene/metal atom/hBN sandwich structures.
 \label{Table2}}%
 \begin{ruledtabular}
 \begin{tabular}{cccccccc}
       &  $U-J$     & Li & K & Cr & Mn & Co & Cu \\
\hline
donated  &  0 eV & 1.00 & 0.80 & 1.18 & 1.16 & 0.66 & 0.67 \\
electrons &  2 eV & $\cdot$ & $\cdot$ & 1.12 & 1.20 & 0.70 & 0.64 \\
         &  4 eV & $\cdot$ & $\cdot$ & 1.05 & 1.16 & 0.70 & 0.61 \\
\hline
spin     & 0 eV  & 0.00 & 0.00 & 3.80 & 3.07 & 1.25 & 0.00 \\
magnetic & 2 eV  & $\cdot$ & $\cdot$ & 4.18 &  3.68 & 1.33 & 0.00 \\
moment   & 4 eV  & $\cdot$ & $\cdot$ & 4.40 &  3.95 & 1.86 & 0.00 \\
 \end{tabular}
\end{ruledtabular}
 \end{table}

To understand the effects of the localization characteristics of the
3$d$ electrons in the transition metal impurity, we consider $U-J$
values of 0, 2, and 4 eV.
As listed in Table II, the electron donations of the 3$d$ transition metal atoms
depend on the $U-J$ values.
As $U-J$ increases, the amount of donated electron {\it decreases} slightlyfor the Cr and Cu atoms.
Electron configurations of Cr and Cu are [Ar]$3d^5$$4s^1$ and [Ar]$3d^{10}$$4s^1$, respectively.
This observation implies that the localization of the 3$d$ electrons prevents the electron donation from the impurity atom to the sheets.
In contrast, the amount of donated electron {\it increases} slightly for the Co atom ([Ar]$3d^7$$4s^2$) with increasing $U-J$ values.
On the other, the Mn atom ([Ar]$3d^5$$4s^2$) does not exhibit any monotonic change with increasing $U-J$ values.
For the spin magnetic moment, all the transition metal atoms that we consider show increasing magnetic moments as $U-J$ increases.

Figure 3 shows the total DOS of the sandwich structures and the PDOS for each value of $U-J$.
For $U-J = 0$ eV, except in the case of the Cu impurity, the 3$d$ states of the
transition metals with majority spin lie between $-2$ and $-1$ eV.
The Mn and Co atoms, in particular, have strong peaks very close to
$E_f$ for the spin-down electron. In addition, the Co atom gives
rise to magnetic impurity states. These results are similar to those
previously reported for Mn and Co adsorption on graphene.\cite{Mao,Johll}
Near the Fermi level, the PDOS of Co originates mainly from the $d_{yz}$ orbital, while that of Mn
originates from the $d_{xz} + d_{yz}$ orbitals. As $U-J$ increases,
the occupied 3$d$ states move deeper inside the valence band. Since the
3$d$ orbitals of the Cu atom are fully occupied, the PDOS of the
occupied 3$d$ states is not shown in the energy range from $-$2 to
$+$2 eV, regardless of the value of $U-J$, although $\sim$0.7 $e$ is
donated to the graphene and hBN sheets. A further increase in $U-J$
does not significantly change the electron transfer, but the spin
magnetic moments of the structures containing Cr, Mn, and Co
increase.

\begin{figure}[tb!]
\includegraphics[width=0.8\columnwidth]{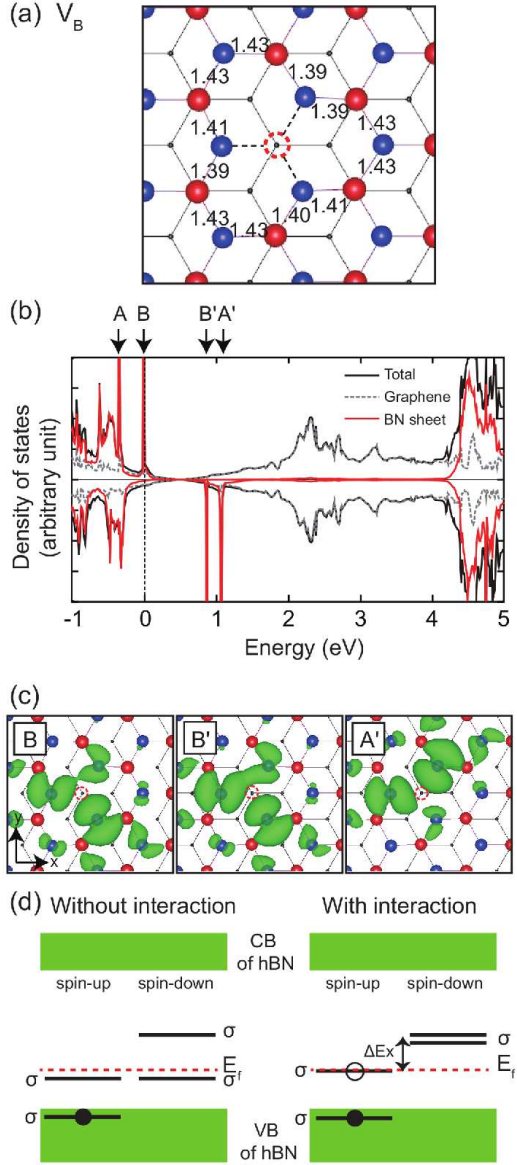}
\caption{(Color online) (a) Optimized structure of a graphene/hBN heterostructure
containing a $V_B$ defect. The red dotted circle
indicates the position of the single boron vacancy. (b) Density of
states of the structure in (a). (c) Partial charge densities for
states B, B$'$, and A$'$. (d) Schematics of the energy levels of the
defect states in the band gap of the hBN sheet induced by $V_B$ in the dilute defect limit.
Filled and empty circles represent fully occupied and partially
occupied levels, respectively, for the spin-up electron. Both the
conduction band (CB) and valence band (VB) have
been labeled; $\sigma$ indicates $\sigma$-bonded electrons.
In (d), $\Delta E_x$ stands for exchange splitting. \label{Fig4}}
\end{figure}

\begin{figure}[tb!]
\includegraphics[width=0.8\columnwidth]{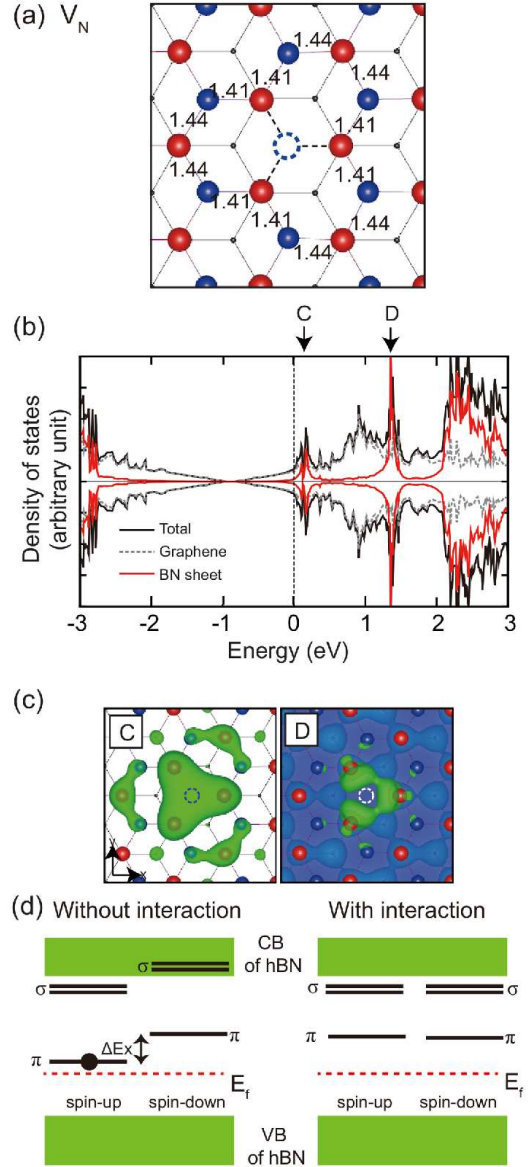}
\caption{(Color online) (a) Optimized structure of a graphene/hBN heterostructure
containing a $V_N$ defect. The blue dotted circle
indicates the position of the single nitrogen vacancy. (b) Density
of states of the structure in (a). (c) Partial charge densities for
states C and D. (d) Schematics of the energy levels of the defect
states in the band gap of the hBN sheet induced by $V_N$. The filled
circle represents a fully occupied level for the spin-up electron.
Both the conduction band (CB) and valence band (VB) have been labeled; $\sigma$ and $\pi$ indicates $\sigma$- and $\pi$-bonded electrons, respectively.
In (d), $\Delta E_x$ stands for exchange splitting. \label{Fig5}}
\end{figure}

\subsection{Intrinsic defects: vacancies in hBN}
Defects within the hBN sheet itself affect the electronic
properties in a different manner. Figure 4(a) shows that when the
structure contains a boron monovacancy ($V_B$), the three neighboring
nitrogen atoms around $V_B$ move slightly to strengthen the bonds
with the neighboring boron atoms. Compared to the B--N bond length
(1.44 \AA) at sites far from $V_B$, those adjacent to $V_B$ are shorter
(1.39--1.41 \AA).

If we assume that the neighboring N atoms form an
equilateral triangle, then their dangling bonds ($\phi_1$, $\phi_2$,
and $\phi_3$) can be weakly coupled and produce three localized
energy eigenstates ($\psi_1$, $\psi_2$, and $\psi_3$).
$\phi$ denotes electronic states of dangling bonds. In spite of the disconnection of $\pi$ electrons of three N atoms,
their energy levels are not much affected by the vacancy. This is not surprising since $\pi$ electrons are localized at the N sites in a perfect hBN.
In the case of the $V_N$ defect, however, the lowest unoccupied level is composed of $\pi$ electrons at the B site.
Thus, both $\pi$ and $\sigma$ electrons should be considered as gap states:
\begin{equation}
\begin{array}{l}
\psi_1 = \frac{1}{\sqrt{3}}\left(\phi_1 + \phi_2 + \phi_3 \right),\\
\psi_2 = \frac{1}{\sqrt{6}}\left(2\phi_1 - \phi_2 - \phi_3 \right),\\
\psi_3 = \frac{1}{\sqrt{2}}\left(\phi_2 - \phi_3 \right).
\end{array}
\end{equation}
Here, $\psi_1$ is the lowest-energy defect state, and $\psi_2$ and
$\psi_3$ are degenerate states with a higher energy.
It has been reported\cite{Azevedo,Attaccalite, Bing1} that the $V_B$ defect
lowers its symmetry from $D_{3h}$ to $C_{2v}$ via the Jahn--Teller
distortion, and therefore, the degeneracy disappears and an energy
splitting of the occupied degenerate defect states occurs.

Figures 4(b) and 4(c) show that $V_B$ produces four localized defect states: A, B, B$'$,
and A$'$. State B is located at $E_f$ but it has $\sigma$-bonding characteristics.
Therefore, this state has negligible hybridization with the graphene states with $\pi$-bonding characteristics.
States B$'$ and A$'$ are around $+$1 eV in the conduction band.
The localized states B and B$'$ would be at the same energy level if there were no interaction between the graphene and hBN sheets.
Because of an exchange splitting of $\sim$1 eV, the spin-degenerate state ($\psi_2$) is separated
into two. Thus, the partial charge densities of state B for the
spin-up electron and state B$'$ for the spin-down electron have
the practically identical shape.
Electron transfer takes place from graphene to hBN, and the Fermi level
decreases in energy.\cite{Si, Azevedo} Our Bader
charge analysis also supports electron donation ($\sim$0.5 $e$) from
graphene to hBN. In fact, the spin magnetic moment of $\sim$1.5
$\mu_B$ induced by $V_B$ is associated with this electron transfer.
According to Huang {\it et al.},\cite{Bing1} the spin magnetic
moment induced by $V_B$ is 1.0 $\mu_B$ in the absence of graphene.
In the absence of the graphene monolayer, we found that there were
one and two unoccupied defect levels of the spin-up and spin-down
electrons, respectively, above the valence band maximum (VBM) of
hBN, which have $\sigma$-bonding characteristics, as shown in Figure 4(d).\cite{Bing2} 
In the presence of the interaction with graphene, the charge ($\sim$0.5 $e$) transferred from graphene increases the spin magnetic moment 
(1.0 $\mu_B$ $\rightarrow$ 1.5 $\mu_B$) in hBN and account for the partially occupied state B and the two unoccupied defect levels 
that correspond to states B$'$ and A$'$ in Figure 4. 
The lowest defect state $\psi_1$ in Eq. (2) is located inside the valence band (state A), 
while $\psi_3$ corresponds to state A$'$.

In the case of a nitrogen monovacancy ($V_N$), the B--N bonds around
$V_B$ have shorter lengths of 1.41 \AA, and in contrast to $V_B$,
$V_N$ does not produce a spin magnetic moment. For a bare hBN sheet, $V_N$ has a
total spin magnetic moment of 1 $\mu_B$.
Since about one electron is donated from hBN to graphene, the Fermi level
increases in energy and graphene becomes $n$-doped.
Interestingly, the electron transfer cancels the spin magnetic moment.
Three localized states can be distinguished clearly in the DOS: one around $E_f$ and two around
1.4 eV that are nearly degenerate. This is reflected in the large
difference in the peaks C and D of the PDOS. Since the $V_N$ defect
preserves the symmetry of $D_{3h}$ and the energy degeneracy is not
broken,\cite{Bing1} state C corresponds to $\psi_1$ and state D to a
linear combination of $\psi_2$ and $\psi_3$.
In the absence of graphene, one spin-up level with $\pi$-bonding characteristics is occupied and one
spin-down level with $\pi$-bonding characteristics is unoccupied
owing to exchange splitting, as shown in Figure 5(d).\cite{Bing2}
However, electron transfer from hBN to graphene results in the
occupied spin-up level becoming unoccupied. These two energy levels
for spin-up and spin-down are related to peak C in Figure 5(b).
Therefore, state C with $\pi$-bonding characteristics hybridizes with the graphene states, as shown in Figure 5(c),
and could influence the electron transport in graphene as a scatterer.

\begin{figure}[tb!]
\includegraphics[width=1.0\columnwidth]{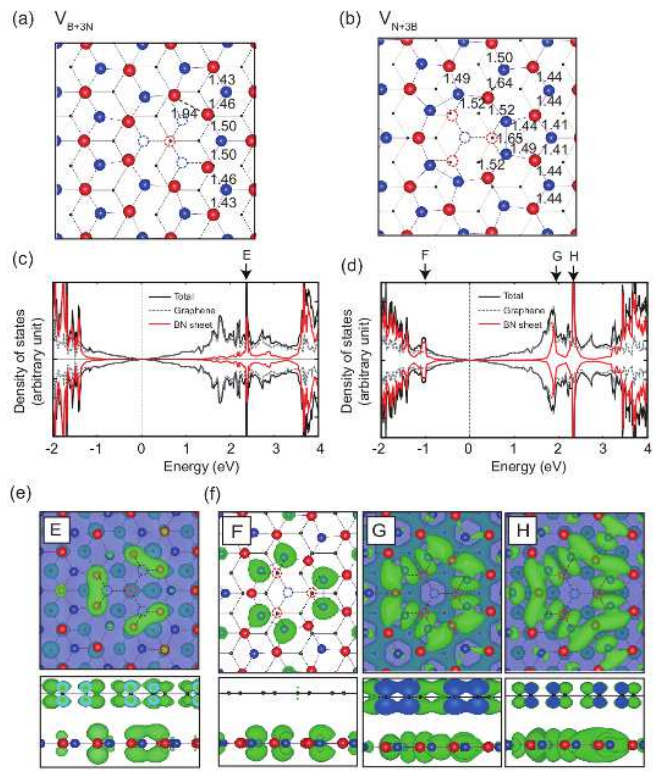}
\caption{(Color online) Optimized structure of a graphene/hBN heterostructure
containing (a) a $V_{B+3N}$ defect and (b) a $V_{B+3N}$
defect. Dotted circles indicate the defect positions. (c) and (d)
Density of states of the structures in (a) and (b), respectively.
(e) Partial charge densities for state E in (c). (f) Partial charge
densities for states F, G, and H in (d). \label{Fig6}}
\end{figure}

We now turn our attention to two smallest triangular defect
structures: $V_{B+3N}$ and $V_{N+3B}$. After geometry optimization,
the distance between two boron atoms in $V_{B+3N}$ decreases to 1.94
\AA~[Figure 6(a)], whereas the surrounding B--N bonds become
slightly weaker and their lengths increase to 1.50 \AA. The
geometrical structure of the hBN sheet in the graphene/hBN heterostructure is very similar to that of
a free-standing hBN sheet in the case of a $V_{B+3N}$ vacancy.\cite{Okada, Yin}
For the $V_{N+3B}$ defect, in contrast, the distance
between two nitrogen atoms decreases to 1.65 \AA, as displayed in Figure 6(b),
and the surrounding B--N bonds become slightly weaker, increasing in length to 1.52 \AA.
The DOS in Figure 6(c) shows that the $V_{B+3N}$ defect creates a localized state, 
corresponding to peak E, deep inside the conduction band, which is a bonding state of the three boron pairs,
as shown in the partial charge density in Figure 6(e).
The $V_{N+3B}$ defect induces one localized state (F) in the valence band and two
localized states (G and H) in the conduction band. State F is an antibonding state of the
three nitrogen pairs. The DOS in Figure 6(d) shows that peak G is
broader than peak H, indicating that state G has a stronger
interaction with the graphene states, which is in agreement with the
partial charge densities in Figure 6(f). Both of our model
structures for the $V_{B+3N}$ and $V_{N+3B}$ defects are
spin-unpolarized, and notably, there are no
dangling bonds in these triangular defects after the geometry
reconstruction. In addition, the defect states (states E to H) are
relatively far from $E_f$, which is in sharp contrast to the
monovacancy cases; $V_B$ and $V_N$ result in the deep levels B and
C, respectively, which are close to $E_f$.

\section{Conclusions}
The electrical properties of graphene  may  be extremely sensitive to the environment.
Therefore, any external contamination easily degrades the conductance of graphene.
Although tremendous efforts have been made to prevent contamination at the interfaces,
it is very difficult to completely remove the contamination.
In the present study, we have studied the influence of
extrinsic defects, such as metal impurities, or
intrinsic defects, such as vacancies, on the structural and electronic properties of the graphene/hBN heterostructure.
We found that a single boron (nitrogen) vacancy in the hBN layer creates
$p$-doped ($n$-doped) graphene and that metal atom impurities may
increase the energy of the Fermi level of graphene. Boron and
nitrogen monovacancies as well as Mn and Co impurity atoms produce
deep levels due to gap states (near the Fermi level).
In contrast, $V_{B+3N}$ and $V_{N+3B}$ defects are expected
to contribute less than monovacancies ($V_B$ and $V_N$) to
electronic scattering in the deposited graphene since their
localized defect states are quite far from $E_f$.
Consequently, the graphene/hBN system with extrinsic contamination or intrinsic vacancy defects could exhibit poor performance
since the imperfections impair the electrical conductivity due to residual scattering in applications such as field effect transistors.

\section*{Acknowledgments}

The authors thank Prof. M. J. Han and Dr. A. T. Lee for helpful discussions.
This work was supported by the Priority Research Center Program (2010-0020207)
and the Basic Science Research Program (2013R1A1A2009131) through the National Research Foundation of the Korea Government.

\end{document}